\documentclass[graybox]{svmult} 
\usepackage[utf8]{inputenc}
 \usepackage{amssymb}
 \usepackage{amsmath}
 \usepackage{mathtools}
\usepackage{amsfonts} 
 \usepackage{setspace}
  \usepackage{mathrsfs}  
\usepackage[normalem]{ulem}

\usepackage{subfig}
\usepackage[usenames,dvipsnames]{xcolor}
\usepackage[pageanchor=false,colorlinks=true,linkcolor=blue,citecolor=blue,urlcolor=blue]{hyperref}
\newcommand\abs[1]{\left|#1\right|}
\newcommand\norm[1]{\left\|#1\right\|}

\newcommand\bE{\mathbb{E}}
\newcommand\bP{\mathbb{P}}

\newcommand\bR{\mathbb{R}}

\newcommand\bRp{\bR_{+}}

\newcommand\sN{\mathbb{N}}

\newcommand\sH{\mathcal{H}}
\newcommand\sA{\mathcal{A}}
\newcommand\sC{\mathcal{C}}
\newcommand\sE{\mathcal{E}}

\newcommand\sS{\mathcal{S}}

\newcommand\sF{\mathcal{F}}

\newcommand\sT{\mathcal{T}}
\newcommand\vx{x}
\newcommand\vy{y}
\newcommand\vz{z}
\newcommand\ve{e}

\newcommand\va{a}

\newcommand\cl[1]{\overline{#1}}
\newcommand{\tam}{\mathrm{argmin}}

\newcommand{\tconv}{\mathrm{conv}}
\newcommand{\tmin}{\mathrm{min}}

\newcommand{\tspan}{\mathrm{span}}

\newcommand{\boxdim}{\mathrm{boxdim}}

\newcommand{\ls}{\langle}
\newcommand{\rs}{\rangle}
\newcommand{\re}{\mathcal{R}e}

\newcommand{\defin}{:=}
\newcommand{\supp}{\textrm{gsupp}}
\newcommand{\linspan}{\textrm{span}}

\newcommand{\mytextbullet}{- }

\author{Yann Traonmilin, Gilles Puy, Rémi Gribonval and Mike E. Davies
}
\institute{Yann Traonmilin  \email{yann.traonmilin@inria.fr}, Rémi Gribonval \at INRIA Rennes - Bretagne Atlantique,
          Campus de Beaulieu 
          35042 Rennes Cedex, France.
\and Gilles Puy \at Technicolor, 975 Avenue des Champs Blancs, 35576 Cesson-Sévigné, France. Gilles Puy contributed to the results reported in this chapter when he was
at INRIA Rennes.
 \and Mike E. Davies \at  Institute for Digital Communications (IDCom), University of Edinburgh,
The King's buildings, Edinburgh, EH9 3JL.}
\date{}
\begin{document}
 \title*{Compressed sensing in Hilbert spaces}
  \maketitle

\abstract{In many linear inverse problems, we want to estimate an unknown vector belonging to a high-dimensional (or infinite-dimensional) space from few linear measurements. To overcome the ill-posed nature of such problems, we use a low-dimension assumption on the unknown vector: it belongs to a low-dimensional model set. The question of whether it is possible to recover such an unknown vector from few measurements then arises. If the answer is yes, it is also important to be able to describe a way to perform such a recovery.
We describe a general framework where appropriately  chosen random measurements guarantee that recovery is possible. We further describe a way to study the performance of recovery methods that consist in the minimization of a regularization function under a data-fit constraint. }
 
 \section{Introduction}

 Many signal processing tasks aim at estimating a signal $x$ from its observation $y$. The signal $x$ can often be described by a continuous physical phenomenon and the observations $y$ are made of a finite collection of scalar measurements. The most basic example of such observations is a sampled version of the signal $x$ (e.g. for a sound recorded at a given sampling rate, the continuous $x$ is the electrical signal produced by the microphone over time). More generally, we consider observations $y$ modeled as 
 \begin{equation}
  y= Ax+e
 \end{equation}
where $x\in \sH$, $y \in \sF$ and $\sH,\sF$ are Hilbert spaces of finite or infinite dimension. The operator $A$ is a linear map and $e$ is a noise whose energy $\|e\|_\sF $ is bounded. In most cases, the operator $A$ models a finite number of measurements $m$. This Hilbert space setting is a way to have a general view of signal recovery problems in classical finite or infinite-dimensional spaces where signals (in a wide sense: time series, images, videos, \ldots) are modelled, e.g. the space of continuous signals with finite energy $\mathcal{L}^2(\bR^d)$, the space of bandlimited signals with finite energy or its equivalent after sampling, $\ell^2(\bR^d)$, or the finite-dimensional vector space $\bR^d$.  

\subsection{Observation model and low complexity signals}

Observing a continuous signal with finitely many linear measurements induces an information loss. If no further prior information on the signal is available, recovering $x$ from $y$ is generally not possible. However, if an (approximate) hypothesis of ``low complexity'' on $x$ is available, enforcing the hypothesis in the recovery process can ensure that we are able to estimate $x$ with reasonable accuracy. Low complexity can be defined in several ways. It often means that the signal lives in a ``low-dimensional model'' or can be described by few parameters. Two classical examples where low complexity helps to recover the signal are: 

\begin{itemize}
 \item Sampling of periodic band-limited signals in $\sH = \mathcal{L}^2(\bR)$: if the signal is known to be band-limited with cut-off frequency $B$, it is possible to recover it perfectly provided it is sampled at a rate at least $2 B$.
 
 \item Compressed sensing in $\sH = \bR^n$: if the signal is known to have at most $k$ non-zero samples in $\bR^n$, it can be recovered with high probability from $m$ random Gaussian (or Fourier) observations provided $m \gtrsim k \log(n)$ \cite{Candes_2008} (We use the symbol $\gtrsim$ to say that there is an absolute constant $C$ such that if $m \geq C k \log(n)$ recovery is possible with high probability). Similarly, if the signal is an  $n \times n$ matrix with rank at most $r$, in the space $\sH = \bR^{n \times n}$, it can be recovered with high probability from $m \gtrsim r n$ random Gaussian observations \cite{Candes_2011a}.
 \end{itemize}
 
In the following, the notion of low complexity is summarized by the fact that $x$ is well approximated by an element of a so-called {\em model set} $\Sigma$, where $\Sigma \subset \sH$ is low-dimensional according to a notion of dimension that will be specified. The considered notion of dimension will be defined in Section~\ref{sec:proj_finite_subspaces} and is related to the number of unknowns we need to estimate to characterize the signal. 
In the context of linear inverse problems with such low-dimensional models, a first objective is to obtain conditions on the linear operator $A$ and the model set $\Sigma$ that guarantee a possible recovery. From this perspective, the analysis of ``low complexity recovery'' is an extension of classical analyses of sparse recovery or low rank matrix recovery.
A second objective, related to the field of compressed sensing, is dimension reduction, where the goal is to design a linear operator $A$ (often with randomness) so that low complexity recovery is possible, with an emphasis on allowing the dimension $m$ of the observation to be small.

 
 \subsection{Decoders}
As the ultimate task is to recover $x$ from $y$, the analysis of the observation of $x$ must be held together with the study of the methods used to recover $x$ from $y$, which we call {\em decoders}. In this chapter, we consider the general class of decoders which consist in minimizing a regularizer under a data fit constraint. We study estimates $x^*$ of $x$ of the form 
\begin{equation} \label{eq:robust_minimization}
   x^* \in \underset{z \in \sH}{\tam} \;   f(z) \;s.t.\;  \|Az-(Ax+\ve)\|_\sF \leq \epsilon.
\end{equation}
Formulation~\eqref{eq:robust_minimization} covers many decoders proposed in the literature, even though other formulations exist (e.g., minimizing $\|Az-(Ax+\ve)\|_\sF$ under a constraint on $f(z)$, or using a Lagrangian formulation). The study presented in this chapter does not require $x^*$ to be the unique minimizer of~\eqref{eq:robust_minimization}. It must be noted that this formulation somehow emphasizes practical signal processing applications because an estimation $\epsilon$ of the observation noise energy $\|e\|_\sF$ is often available (e.g. in photography the noise level can be estimated using aperture time and light conditions, in image processing more advanced techniques allow to estimate noise level from an image \cite{Sutour_2015}). 
 
 The main parameter of the decoder is the regularizer $f$. Its role is to force the estimate belongs to the chosen model set. The form of the data fit constraint  ($\|\cdot\|_\sF$) influences the types of noise that the decoder can robustly manage. This raises interesting questions that are, however, out of the scope of this chapter. The main qualities required for a decoder are: 1) to provide {\em exact recovery of vectors $x \in \Sigma$ in the noiseless setting}; 2) to be \emph{stable to observation noise and robust to modeling error}. 
 
 We emphasize the role of two classes of decoders: ``ideal'' decoders and convex decoders. 
 \begin{itemize}
  \item  Given a problem with a model set $\Sigma$, the \emph{ideal decoder} corresponds to minimizing \eqref{eq:robust_minimization} using $f := \iota_\Sigma$ the characteristic function of $\Sigma$, i.e. $\iota_\Sigma(x) = 0$ if $x \in \Sigma$, $\iota_\Sigma(x) = \infty$  otherwise. This decoder is called ideal, as it enforces perfectly the fact that the solution must belong to $\Sigma$ (the prior on the unknown). Unfortunately, it is generally hard to calculate efficiently as the function to minimize is both non-convex and non-smooth. Consequently, we often use a heuristic for the minimization or turn to a convex proxy to this minimization.
  
  \item The decoder is said to be a \emph{convex decoder} when $f$ is convex. Such a decoder is often easier to compute as the minimization problem has no local minimum other than the global minima even if it this does not guarantee that the minimization can be efficiently performed, see e.g. tensor recovery problems~\cite{Hillar_2013}. State of the art shows that having some additional hypothesis on the linear operator $A$ enables to guarantee stability and robustness of certain convex decoders for classical model sets $\Sigma$.
 \end{itemize}

 \subsection{The RIP: a tool for the study of signal recovery}
 As we just saw, studying signal recovery amounts to studying the interactions between the model $\Sigma$, the regularization $f$ and the measurement operator $A$. We propose here to use a tool that enables us to separate the study of $A$ with respect to $\Sigma$ from the study of $f$ with respect to $\Sigma$: the restricted isometry property (RIP). It is generally defined in our setting for a linear observation operator $A$ on the so-called \emph{secant set} $\Sigma-\Sigma := \{x-x': x\in\Sigma,x'\in\Sigma\}$ 
\begin{definition}[RIP]
The linear operator $A: \sH \to \sF$ satisfies the RIP on the secant set $\Sigma-\Sigma$ with constant $\delta$ if for all $\vx \in \Sigma-\Sigma$:
\begin{equation}\label{eq:DefRIP}
(1-\delta)\|\vx\|_\sH^2 \leq \|A \vx\|_\sF^2 \leq (1+\delta)\|\vx\|_\sH^2
\end{equation}
where $\|\cdot\|_{\sH}$ and $\|\cdot\|_{\sF}$ are Euclidean norms on $\sH$ and $\sF$. 
\end{definition}

\begin{figure}[ht]
\centering
\subfloat{\includegraphics[width=0.75\linewidth]{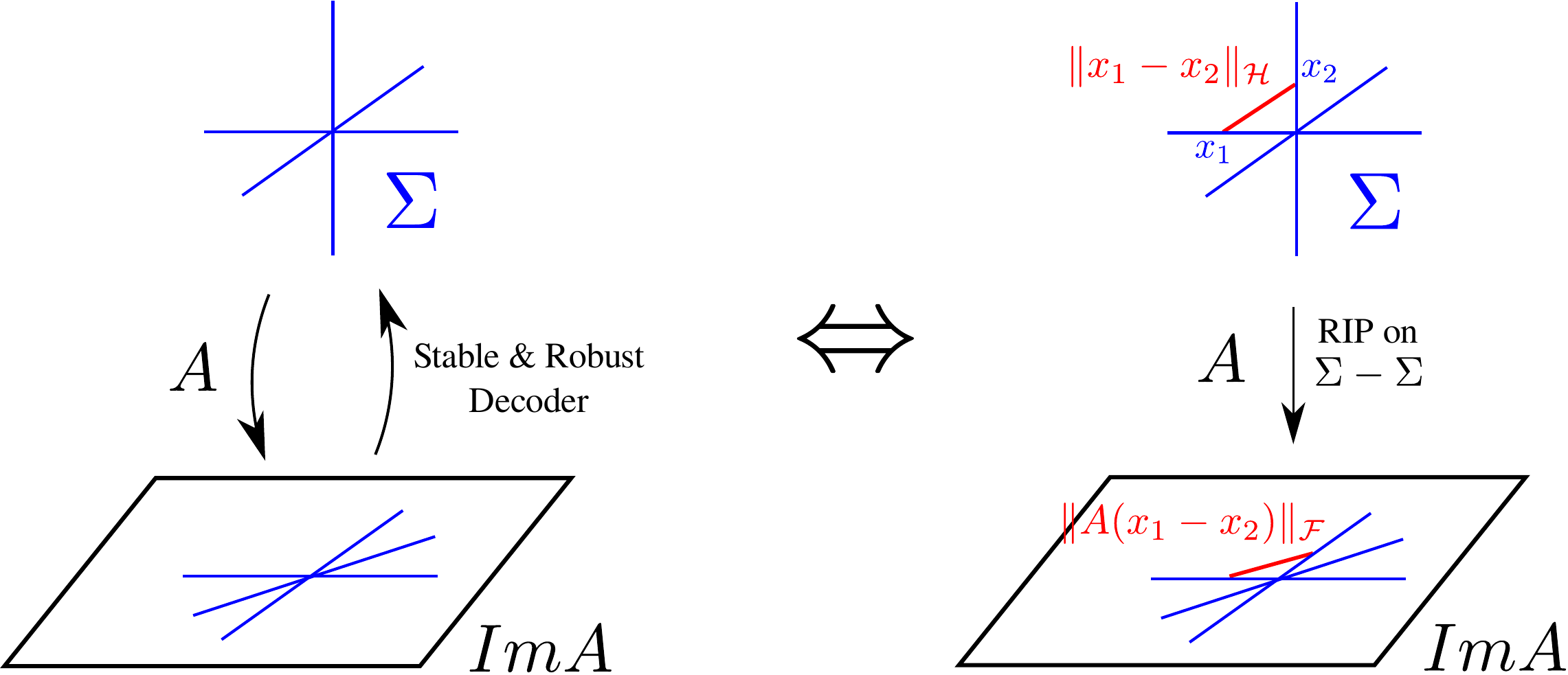}}
\caption{A graphical representation of the equivalence between the existence of stable robust decoders and the RIP on the secant set. Operators satisfying the RIP approximately preserve distances between elements of $\Sigma$.}\label{fig:RIP}
\end{figure}

This property is a famous sufficient condition on $A$ to guarantee the success of convex decoders~\eqref{eq:robust_minimization} in the case of sparse and low rank signal recovery for appropriately chosen regularization $f$~\cite{Donoho_2006,Candes_2006b,Recht_2010, Candes_2010, Chandrasekaran_2012,Foucart_2013}. Intuitively, the RIP requires the operator $A$ to preserve the distance between any two elements of $\Sigma$ (see Figure~\ref{fig:RIP}). 
Moreover, a lower RIP is a necessary condition for the existence of stable and robust decoders: given $A$ and $\Sigma$, if a stable and robust decoder exists then, up to a global rescaling, $A$ satisfies a lower RIP on the secant set $\Sigma-\Sigma$~\cite{Cohen_2009,Bourrier_2014}. 

For example, in the case of sparse recovery, it is possible to show two facts. 
\begin{itemize}
 \item Fact 1: Random Gaussian matrices of size $m \times n$ satisfy the RIP on the set of $2k$-sparse vectors (the secant set of the set of $k$-sparse vectors) with constant $\delta < 1$ with high probability, provided $m \gtrsim \delta^{-2} k \log(n)$.  
 \item Fact 2: As soon as $A$ satisfies this RIP with constant  $\delta < 1/\sqrt{2}$, it is guaranteed that minimization~\eqref{eq:robust_minimization} with $f(\cdot) =\|\cdot\|_1$, the $\ell^1$ norm, yields stable and robust recovery of all $k$-sparse vectors \cite{Cai_2014}.
\end{itemize}
We see that the study of recovery guarantees in this case is separated in two steps: 1) a study of the behaviour of the linear operator $A$ with respect to the model set $\Sigma$ (in terms of RIP property); and 2) a study of the behaviour of the regularizer $f$ with respect to the model set $\Sigma$, that has consequences for all operators satisfying a RIP with a small enough constant. 

The framework presented in the following generalizes these features in order to manage not only the classical sparse recovery/low-rank recovery and related compressed sensing theory, but much beyond to many sorts of low-dimensional model sets.

 \subsection{A general compressed sensing framework}
 
\begin{figure}
\centering
\includegraphics[width=0.4\columnwidth]{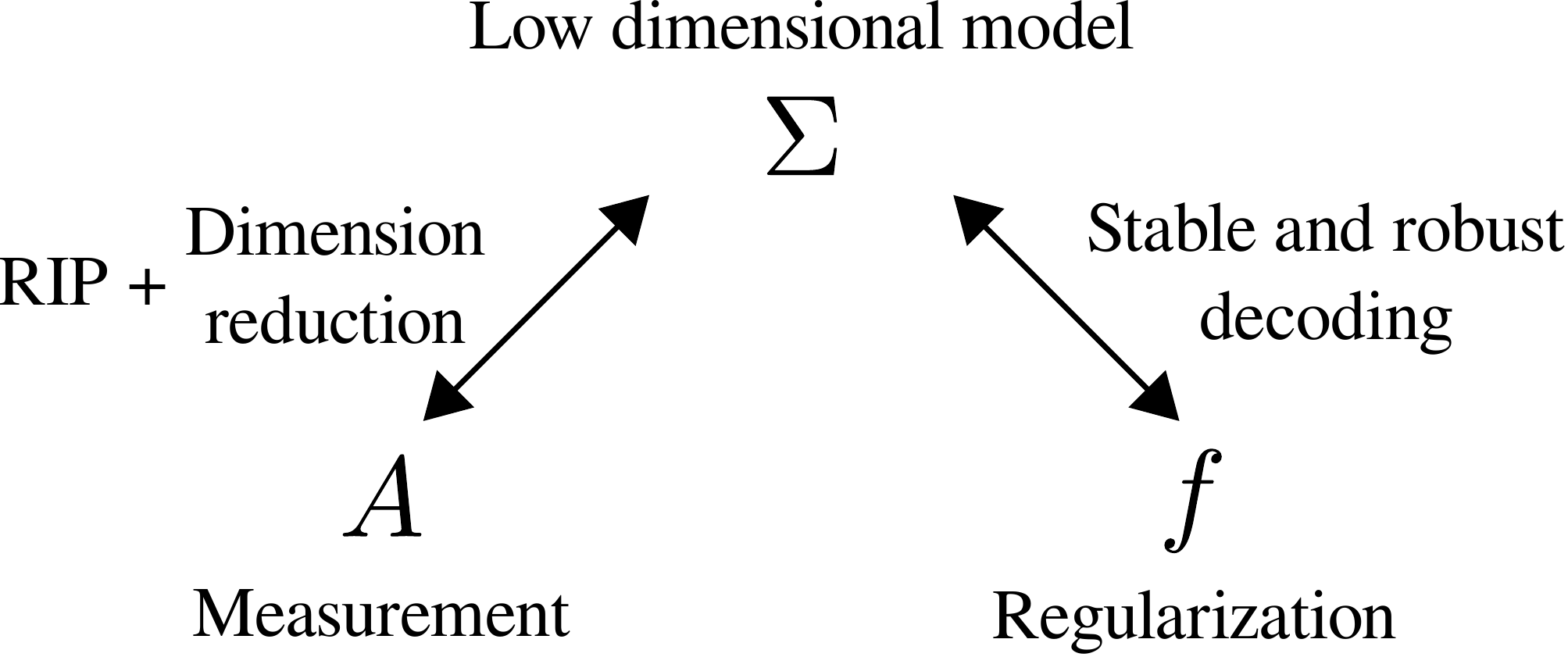}
\caption{Structure of the framework: the RIP framework allows to separate the study of dimension reduction and of decoding.}\label{fig:outline}
\end{figure}

 The remaining part of this chapter shows how it is possible to generalize the  steps we just mentioned. The proposed framework\footnote{This chapter gives a unified view of the latest developments in the area found in \cite{Puy_2015} and \cite{Traonmilin_2016}. } consists in answering the following questions (summarized in Figure~\ref{fig:outline}): 
 \begin{itemize}
  \item Low-dimensional model: when is $\Sigma$ ``low-dimensional''? (Section~\ref{sec:low_dimension})
  \item Dimension-reduction: given $\Sigma$, is there an operator $A$ that satisfies the RIP on $\Sigma-\Sigma$? What level of dimension reduction can it achieve?  (Section~\ref{sec:observing})
  \item What is a good regularizer? Given $\Sigma$ and $f$, does a RIP of $A$ on $\Sigma-\Sigma$ guarantee that $f$ recovers the elements of $\Sigma$?  (Section~\ref{sec:recovering})
 \end{itemize}
 
Section~\ref{sec:generality} mentions generalizations that were left out of the main of the chapter in order to keep the exposition accessible, 
and discusses what challenges we face to go beyond this general compressed sensing framework in Hilbert spaces.

 \section{Low-dimensional models} \label{sec:low_dimension}
 
We begin by precisely describing the low-dimensional models that will be considered in this chapter. We then focus on a model of {\em structured sparsity in levels}, which we will use as a running example to  illustrate the different concepts used in this chapter.
 \subsection{Definition and examples}
\begin{figure}
\centering
\includegraphics[width=0.8\columnwidth]{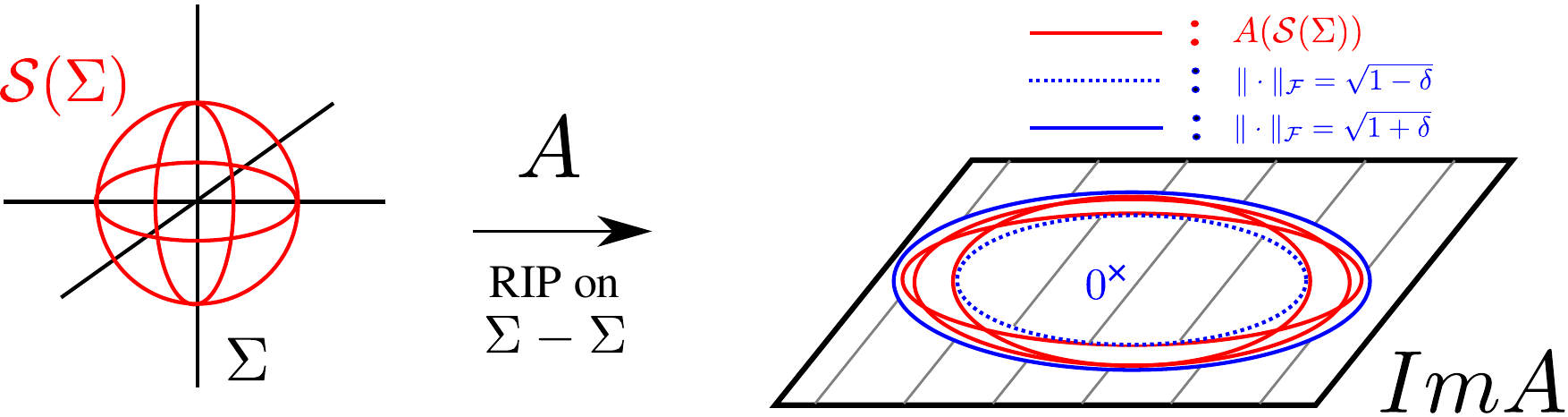}
\caption{A characterization of the RIP: the RIP on the normalized secant set. The image of the secant set must lie within a distance $\delta$ of the unit sphere.}\label{fig:normalized_secant}
\end{figure}

The results presented in \cite{Puy_2015} show that one can always construct a linear operator $A$ that satisfies the RIP on $\Sigma - \Sigma$ if its normalized secant set $\mathcal{S}({\Sigma})$ has a finite intrinsic dimension. The \emph{normalized secant set} of $\Sigma$ is defined as 

\begin{align*}
\mathcal{S}({\Sigma}) := \left\{ z = \frac{y}{\|{y}\|_{\sH}}: y \in (\Sigma-\Sigma) \setminus \{\vec{0}\} \right\}.
\end{align*}
 We substitute $\mathcal{S}$ for $\mathcal{S}({\Sigma})$ hereafter to simplify notations. We illustrate in Figure~\ref{fig:normalized_secant} the RIP on the normalized secant set which is equivalent to the RIP on the secant set. 
 
In this chapter, we measure the intrinsic dimension of $\mathcal{S}$ using the upper box-counting dimension, which is linked to the notion of covering number. 

\begin{definition}[Covering number]
Let $\alpha >0$ and $\sS \subset \sH$. The covering number $N(\sS,\alpha)$ of $\sS$ is the minimum number of closed balls  (with respect to the norm $\| \cdot \|_\sH$) of radius $\alpha$, with centers in $\sS$, needed to cover $\sS$.
\end{definition} 

The upper box-counting dimension is then defined as follows.

\begin{definition}[Upper box-counting dimension]
The upper box-counting dimension of $\sS$ is

\begin{equation*}
\boxdim (\sS) :=  \limsup_{\alpha \rightarrow 0} \; {\log[N(\sS,\alpha)]}/{\log[1/\alpha]}.
\end{equation*}
\end{definition}
Hence, as soon as $k > \boxdim(\sS)$, there exists a model-set dependent constant $\alpha_{\sS}~\in~(0, 1/2)$ such that $N(\sS,\alpha) \leq \alpha^{-k}$ for all $\alpha \leq \alpha_{\sS}$. Further, if the covering number satisfies
\begin{equation}
N(\sS,\alpha) \leq \left(\frac{C}{\alpha}\right)^k
\end{equation}
then $\boxdim(\sS) \leq k$.
 
We choose this definition of intrinsic dimension for two reasons. First, for many useful signal models -- \emph{e.g.}, sparse vectors, low-rank matrices, smooth manifolds -- the upper box-counting dimension of the normalized secant set is known. The results presented in this chapter can thus be directly applied to these sets, without additional work. Second, one should be careful with the definition of intrinsic dimension used in an infinite-dimensional space. Indeed, for some definitions of dimension, there are examples where it is impossible to perform dimension reduction on vectors belonging to a set having a finite dimension (i.e. the set cannot be linearly and stably embedded in a finite-dimensional space~\cite[Chapter 6.1]{Robinson_2010}). The upper box-counting dimension of the normalised secant set does not suffer from this issue.
 
 In the following we will say informally that a model $\Sigma$ is \emph{low-dimensional} if $\boxdim (\sS(\Sigma))$ is small compared to the ambient dimension of the Hilbert space $\sH$ (which may be infinite).  In many examples, the dimension $\boxdim (\sS)$ is of the order of the number of parameters needed to describe elements of the model, as in the case of classical sparsity or low rank matrices. For $k$-sparse vectors, the dimension of the normalized secant set $\sS$ is of the order of $k$ \cite[Section C.2]{Foucart_2013}.  For $n \times n$  matrices of rank lower than $r$, the dimension of the normalized secant set $\sS$ is of the order of $rn$. 
 
\subsection{Structured sparsity ...}
As a running example, we use a refinement of the notion of sparsity as a way to introduce the general framework: we consider a model of {\em structured sparsity in levels}.  

We start by describing {\em structured sparsity}, a now classical generalization of the plain sparsity model. In many applications, signals are not only sparse but also clustered in groups of significant coefficients in a transformed domain (Fourier domain, Radon domain,...). Structured sparsity (also called group-sparsity) is the assumption that the signal is supported on a few groups of coefficients~\cite{Gribonval_2008,Baraniuk_2010,Eldar_2010}. 

Formally, we consider an orthonormal Hilbert basis $(\ve_i)_{i \in \sN}$ of $\sH$ and a finite collection  $G$ of non-overlapping finite {\em groups} of indices, i.e. subsets $g \subset \sN$ with $|g| < \infty$ and $g \cap g' = \emptyset$ whenever $g \neq g'$. The restriction of the vector $\vx \in \sH$ to the group $g$ is $\vx_{g} \defin \sum_{i \in g} \ls \vx,\ve_{i} \rs \ve_{i}$. A {\em group support} is a subset $T \subset G$ and the restriction of $\vx$ to the group support $T$ is $\vx_T \defin \sum_{g \in T} \vx_{g}$. The group support of $\vx \in \sH$, denoted $\supp(\vx)$, is the smallest $T \subset G$ such that $\vx_T=\vx$. The size of the group support of $\vx$, denoted $|\supp(\vx)|$, is the cardinality of $\supp(\vx)$ (to be distinguished from the number of non zero coordinates in $\vx$).

Given an integer $k$, the {\em $k$-group-sparse model} is defined as
\begin{equation}\label{eq:DefKGroupSparseModel}
\Sigma_k \defin \{\vx \in \sH,\ |\supp(\vx)| \leq k\}.
\end{equation}

Let $d$ be the size of the biggest group. We have the following covering of $\sS (\Sigma_k)$: 
\begin{equation}
 N(\sS (\Sigma_k), \alpha) \leq \left(\tfrac{C}{\alpha} \right)^{dk}\\   
\end{equation}
where $C$ is a constant depending on $d$.

\subsection{... in levels}
\begin{figure}
\centering
\includegraphics[width=0.9\columnwidth]{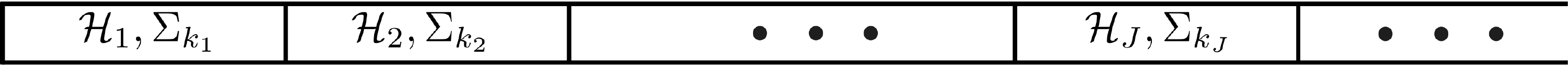}
\caption{A representation of structured sparsity in levels in $\sH$. A structured sparsity in level model is formed by different structured sparsity models in orthogonal subspaces.}\label{fig:sparsity_in_levels} 
\end{figure}

 Consider a collection of $J$ 
 orthogonal spaces $\sH_j \subset \sH$ each equipped with a $k_j$-group-sparse model $\Sigma_j$ as defined in~\eqref{eq:DefKGroupSparseModel} (each with its Hilbert basis and its set $G_j$ of groups). Since the subspaces are orthogonal, there is a natural isomorphism between their direct sum and their Cartesian product. It is simpler to work with the latter, and {\em structured sparsity in levels} is associated to the model (see Figure~\ref{fig:sparsity_in_levels})
\begin{equation}\label{eq:DefStructuredSparseLevelsModel}
\Sigma \defin \left\{\vx \in \sH, \vx = \sum_{j=1}^{J} \vx_{j}, \vx_{j} \in \Sigma_{k_j}\right\},
\end{equation}
which is identified to the Cartesian product of the models $\Sigma_{k_1} \times \Sigma_{k_2} \times \ldots \times \Sigma_{k_j}$. 

Two examples were this model is useful are: medical imaging (MRI) and simultaneous signal and noise sparse modeling~\cite{Adcock_2013b,Studer_2013, Traonmilin_2015}:
\begin{itemize}
 \item In MRI, the different levels where the signal is sparse are wavelet scales.  MRI images are generally sparser at fine wavelet scales than large wavelet scales. This allows for more flexibility in the modeling of the signal than the simple sparsity model. 
 \item Simultaneous signal and noise sparse modeling is a convenient setting for the separation of a signal sparse in some domain from noise that is sparse in another domain.  An observed signal $y$ is modeled as the super-imposition of two components, $y = A_{1}x_{1}+A_{2}x_{2}  $ where $A_{1}x_{1}$ is the signal of interest, $x_1$ lives in the (structured) sparse model $\Sigma_{k_1}$, $A_{2}x_{2}$ is noise, and $x_2$ lives in the (structured)  sparse model $\Sigma_{k_2}$. This model is also related to the separation of transients from stationary parts in audio, or for the decomposition of images into cartoon and texture~\cite{Kutyniok_2013}.  As $y   = [A_{1}\ A_{2}]x$ with $x = [x_{1}^{T},\ x_{2}^{T}]^{T}$, this corresponds to a  two-level (structured) sparse model for $x$.
\end{itemize}
 For structured sparsity in levels, we have \cite{Traonmilin_2016}: 
\begin{equation}
\begin{split}
N(\sS, \alpha) &\leq  N(\sS(\Sigma_{k_1}, \alpha)\times \ldots \times N(\sS(\Sigma_{k_j}), \alpha)\\
&\leq \left(\tfrac{C_1}{\alpha} \right)^{d_1k_1} \times \ldots \times \left(\tfrac{C_J}{\alpha} \right)^{d_J k_J}  \\   
\end{split}
\end{equation}
where $C_j$ are constants that are of the order of the dimension of each level times the maximum size of groups $d_j$  in level $j$. Hence up to log factors, the upper box-counting dimension of $\sS$ in this case is of the order of $\sum  d_jk_j$.

 \section{Dimension reduction with random linear operators} \label{sec:observing}
 Now that we have defined the notion of dimension of a model $\Sigma$ that we work with, and the desirable RIP property of a linear operator $A$, the remaining question is: how to construct a {\em dimension-reducing} linear operator $A: \sH \rightarrow \mathbb{R}^m$ that satisfies the RIP on $\Sigma-\Sigma$? 
 
 Consider an MRI-like scenario with a sparsity in levels signal model $\Sigma_{k_1} \times \ldots \times \Sigma_{k_J}$ in a wavelet basis.  The fact that the signals in $\Sigma$ have a support restricted to the first $J$ wavelet scales implies that their energy decreases at high frequencies. Intuitively, it thus seems unnecessary to probe very high frequencies in the measurement process for this type of signals~\cite{Adcock_2016}. A good approximation of the signals can be obtained by probing all frequencies up to a certain bandlimit $B$. This process corresponds to a projection from the infinite-dimensional space $\sH$ to a finite-dimensional space of size $B$. However, the dimension $B$, though finite, might still be reduced. Indeed, the signals are not just concentrated in the first $J$ wavelet scales, they are also sparse in levels. A dimension-reducing step can thus be envisioned after the projection onto the first $B$ Fourier coefficients with, \emph{e.g.}, a random Gaussian matrix. Ideally the final dimension $m$ should satisfy, up to log factors, $m = O\left(\sum_{j=1}^J k_j\right)$ (of the order of the number of parameters describing the model). Intuition thus suggests to build the operator $A$ in two steps: a projection onto a finite (but high) dimensional space followed by a multiplication with a random matrix. 

\begin{figure}[h]
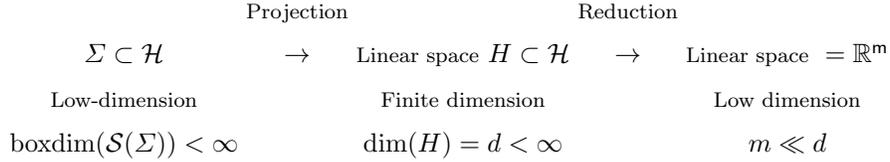

\centering
\begin{tabular}{ccccc} 
& {\small Projection} &  &{\small Reduction}& \\

 $\Sigma \subset \sH$ & $\rightarrow$ &{\small Linear space} $H \subset \sH$  &$\rightarrow$& {\small Linear space} $\sf=\bR^m$ \\
  {\small Low-dimension}&  &{\small Finite dimension}&  & {\small Low dimension} \\
    $\boxdim(\sS(\Sigma))<\infty$ &  &  $\dim(H)=d<\infty$ &  & $m \ll d$ \\

\end{tabular}

\caption{Strategy for dimension reduction: we aim at reducing the dimension of vectors belonging to $\Sigma$ leaving an infinite-dimensional space $\sH$.}\label{fig:dimension_reduction}
\end{figure}

In fact, the authors of \cite{Puy_2015} present such a construction in the general setting  that first projects the signal onto a subspace $H \subset \sH$ of finite (but potentially large) dimension, then reduces the dimension using a random linear operator on $H$ (see Figure~\ref{fig:dimension_reduction}).
 
 \subsection{Projection on a finite-dimensional subspace}\label{sec:proj_finite_subspaces}


Assuming that $\boxdim(\sS)$ is finite we will see that, given $0<\alpha<1$,  there always exists a finite-dimensional subspace $H \subset \sH$ such that 
\begin{equation}
\label{eq:projection}
(1-\alpha) \|x\|_\sH \leq \|P_H x\|_\sH \leq \|x\|_\sH
\end{equation}
for all $x \in \Sigma - \Sigma$, where $P_H$ denotes the orthogonal projection onto $H$.

\begin{figure}
\centering
\includegraphics[width=\columnwidth]{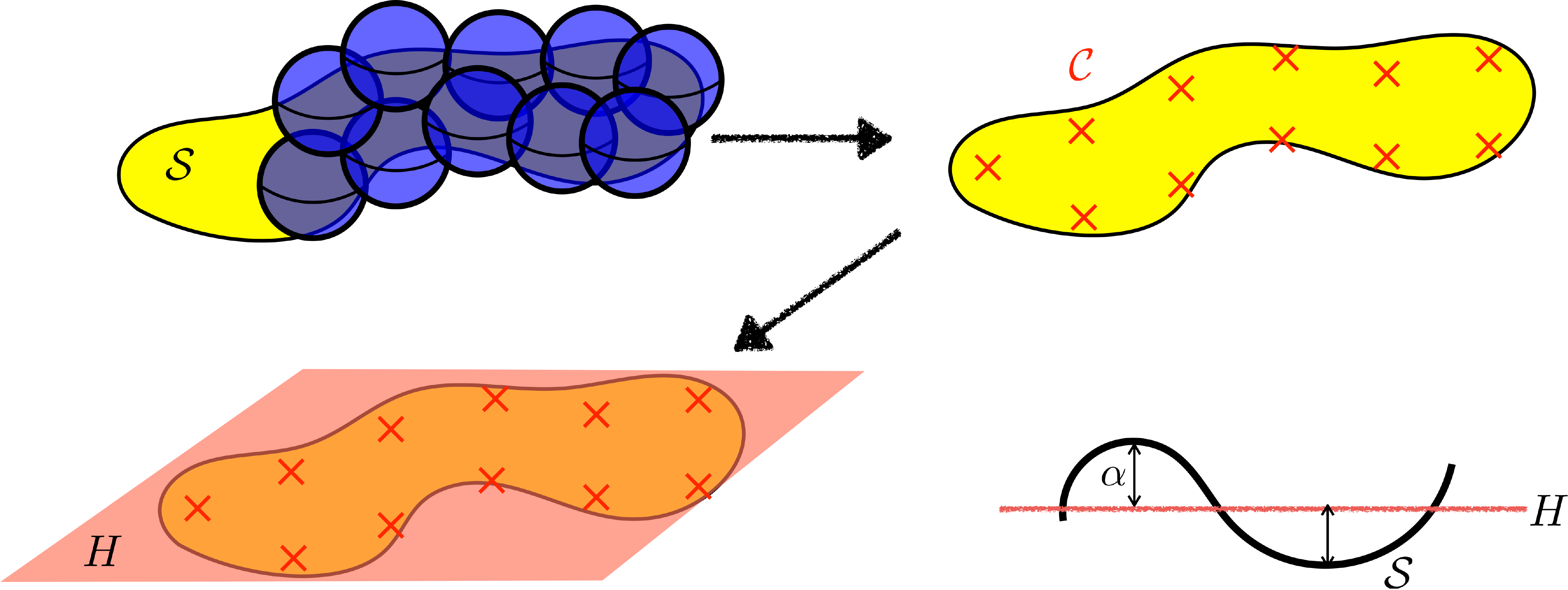}
\caption{\label{fig:proj_H} Construction of $H$. \emph{Top left}: cover of $\sS$ with $N(\sS,\alpha)$ balls of radius $\alpha$. \emph{Top right}: the centers of the balls, indicated by the red crosses, form an $\alpha$-cover, denoted by $\sC$, for $\sS$. \emph{Bottom left}: $H$ is defined as the linear span of the vectors in $\sC$. \emph{Bottom right}: $H$ approximates $\sS$ with precision $\alpha$.}
\end{figure}

In the example of Fourier sampling of signals sparse in a Haar basis, it is possible to directly exhibit such a projection $P_H$ by sampling low Fourier frequencies. However, one can generally construct $H$ as follows. First, build an $\alpha$-cover of the normalized secant set $\sS$. As $\boxdim(\sS)$ is finite, $N(\alpha, \sS) < + \infty$ balls are sufficient to build this cover. Let now $\sC$ be the set containing the center of these balls. It is then sufficient to take $H= \tspan \;\sC$, see Figure~\ref{fig:proj_H} and \cite{Puy_2015}. We remark that in the worst case the cardinality of $\sC$ is exponential in the dimension of $\sS$, hence $H$ can have a dimension of the order of $e^{c(\alpha) \times \boxdim(\sS)}$. Yet the important message to take away at this stage is that: 

\begin{center}
\emph{If the normalized secant set $\sS(\Sigma)$ has a finite upper box-counting dimension, then there exists a finite-dimensional subspace $H \subset \sH$ that approximates all vectors in $\sS$ with precision $\alpha$.}
\end{center}

In the next section, we describe how to further reduce the dimension to $m = O(\boxdim(\sS))$ after this first projection.

\subsection{Dimension reduction step}
  
After the projection onto the finite-dimensional space $H$ of the previous section, the goal is now to reduce the dimension down to $O(\boxdim(\sS))$. As most compressive sensing techniques use random observations to reduce the dimension, it seems natural to follow this route. 

Denote by $d$ the dimension of the subspace $H$ and $(e_1, \ldots, e_d)$ an arbitrary orthonormal basis of $H$. By abuse of notation, identify the projection onto $H$ with the linear operator $P_H: \sH \rightarrow \mathbb{R}^d$ that returns the coordinates of the orthogonal projection onto $H$ in the basis $(e_1, \ldots, e_d)$. The idea is now to compose $P_H$ with a random matrix $M \in \mathbb{R}^{m \times d}$ to build $A: \sH \rightarrow \mathbb{R}^m$, \emph{i.e.}, $A = M P_H$.  Ideally, we would like $A$ to satisfy the RIP, and $m \approx O(\boxdim(\sS))$: a number of measurements of the order of the dimension of the model. In this case, we would be assured that the ideal decoder is stable and robust and that the reduction of dimension is close to optimal. 

\subsubsection{Randomized dimension reduction}

To exhibit a linear operator $A$ satisfying the RIP with constant $\delta$, one can first identify a finite-dimensional subspace $H \subset \sH$ such that~\eqref{eq:projection} holds with $\alpha$ small enough, then {\em build} a random $M: \bR^{d} \to \bR^{m}$ satisfying a RIP with small enough constant $\delta'$. Sometimes one is directly provided with a random linear operator from $\sH \to \bR^{m}$ and needs to {\em check} whether the RIP holds with high probability. The approach described in~\cite{Puy_2015} makes it possible to handle both cases. With a slight abuse of notation, in this subsection $\sH$ stands either for the original Hilbert space (case of a given random operator) or for $\bR^{d}$ (two-step construction considered above).

Consider $M$ a {\em random} linear operator from $\sH$ to $\bR^{m}$. An example results from the independent draw of $m$  identically distributed random vectors $\va_{i} \in \sH$, so that for $\vx \in \sH$, $M\vx := (\langle \va_{i},\vx\rangle)_{i=1}^{m}$. A convenient way to help $M$ satisfy the RIP is to choose its probability distribution so that, for any vector $\vx \in \sH$, 
\begin{equation}
\label{eq:Isotropy}
\bE_{M} \|Mx\|_2^2 = \|x\|_\sH^2. 
\end{equation}
With the above {\em isotropy assumption}, a draw $M$ of the random linear operator satisfies the RIP on $\Sigma - \Sigma$ if, and only if,
\begin{equation}
\label{eq:RIP3}
\left|\|M x\|_2^2 - \bE_{\tilde{M}}\|\tilde{M}x\|_2^2\right| \leq \delta \|x\|_\sH^2,
\end{equation}
for all $x \in \Sigma - \Sigma$, where we emphasize with the $\tilde{M}$ notation that the expectation $\bE$ is with respect to a linear operator with the same {\em distribution} as the one from which the particular $M$ is drawn. As discussed in Section~\ref{sec:generality}, even without the isotropy assumption~\eqref{eq:Isotropy}, one can establish dimension reduction results using~\eqref{eq:RIP3}  as a generalized definition of the RIP \cite{Puy_2015}.

  
To prove that $M$ satisfies the RIP, the authors in \cite{Puy_2015} require it to satisfy two concentration inequalities. Define
\begin{equation*}
\begin{split}
 h_{M} \colon \sH 	& \longrightarrow  	\bR  \\
 x 	& \longmapsto 	\|Mx\|_2^2 - \|x\|_\sH^2.
 \end{split}
\end{equation*}
The assumption is that there exists two constants $c_1, c_2 \in (0, \infty]$ such that for any fixed $y, z \in \sS(\Sigma) \cup \{ 0 \}$,
\begin{equation}
\label{eq:prob_bound_increment_1}
\bP_{M} \left\{ \abs{h_{M}(y) - h_{M}(z)} \geq \lambda \norm{y - z}_\sH \right\}
\leq 2 e^{- c_1 m \lambda^2},
\quad\text{for}\ 0 \leq \lambda \leq {c_2}/{c_1}
\end{equation}
%
%
\begin{equation}
\label{eq:prob_bound_increment_2}
\bP_{M} \left\{ \abs{h_{M}(y) - h_{M}(z)} \geq \lambda \norm{y - z}_\sH \right\}
\leq 2  e^{- c_2 m \lambda},
\quad\text{for}\ \lambda \geq {c_2}/{c_1}.
\end{equation}
%
By taking $z = 0$ in \eqref{eq:prob_bound_increment_1} and \eqref{eq:prob_bound_increment_2}, we see that the above properties imply that, for any fixed vector in the normalized secant set, $y \in \sS$, $\|My\|_2^2$ stays close to its expected value $\bE_{M}\|My\|_2^2 = \|y\|_\sH^{2} =1$ with high probability. Proving that the RIP holds consists in showing that, with high probability on the draw of $M$, this property actually holds uniformly for \emph{all} vectors in $\sS$, not just for any fixed vector $y \in \sS$. Among other properties, this generalisation to the entire set $\sS$ is proved by using the fact that for any fixed $y, z \in \sS$, if $\norm{y - z}_\sH$ is small then the difference between $\|My\|_2^2 - \|Mz\|_2^2$ and $\bE_{\tilde{M}}\|\tilde{M}y\|_2^2 - \bE_{\tilde{M}}\|\tilde{M}z\|_2^2$ is also small with high probability.

These concentration inequalities together with the finite dimension of $\sS$ suffice to conclude on a sufficient number of measurements for $M$ to satisfy the RIP \cite[Theorem II.2]{Puy_2015}.

\begin{theorem}
\label{th:main_theorem}
Let $M~:~\sH \rightarrow \bR^m$ be a random linear map that satisfies \eqref{eq:prob_bound_increment_1} and \eqref{eq:prob_bound_increment_2}. Assume that $\boxdim(\sS)<s$ (there exists $0<\alpha_\sS<\frac{1}{2}$ such that $N(\sS,\alpha) \leq \alpha^{-s}$ for all $0<\alpha < \alpha_{\sS}$). 

Then for any $\xi, \delta_0 \in (0, 1)$, $M$ satisfies the RIP on $\Sigma - \Sigma$ with constant $\delta \leq \delta_0$ with probability at least $1 - \xi$ provided that
\begin{equation}
\label{eq:cond_m_main_theorem}
m \geq \frac{1}{\delta_{0}^2}\frac{C}{\min(c_1, c_2) } \, \max\left\{s \log\left(\frac{1}{\alpha_{\sS}}\right),\log\left(\frac{6}{\xi}\right)\right\},
\end{equation}
where $C>0$ is an absolute constant.
\end{theorem}
  
This theorem states that if the random operator $M$ satisfies appropriate concentration inequalities and the set $\sS$ has finite upper box-counting dimension, then reducing the dimension of the vectors in $\Sigma$ is possible (recall that these vectors possibly live in an infinite-dimensional space). A number of measurements $m$ of the order of the dimension of the secant set $\sS$ (only) is sufficient to be able to recover elements of $\Sigma$, whatever the ambient dimension of $\sH$ (which can be infinite).  We remark that the sufficient number of measurements grows as the RIP constant decreases (the closer $A$ is to an isometry for elements in $\sS$). In particular, the typical $\log n$ factor appearing in standard results for compressed sensing of $k$-sparse vectors in $\sH = \bR^{n}$ is in fact related to $\alpha_{\sS}$ rather than the ambient dimension. Related results, independent of the ambient dimension, have been achieved for manifold embedding \cite{Eftekhari_2015,Dirksen_2014}.

For a fixed dimension of $\sS$, if we wish to ensure an arbitrarily small probability that the RIP fails to hold,  $\xi \leq 6 \left(\alpha_\sS\right)^s$, then the number of measurements $m$ also grows as $\xi$ approaches zero. Vice-versa, as the ratio between $m$ and its minimum value $m_{0} = \frac{1}{\delta^2}\frac{C}{\min(c_1, c_2) } \, s \log\left(\frac{1}{\alpha_{\sS}}\right)$ grows, the RIP holds with probability exponentially close to $1$. 

\begin{remark}
The sufficient condition $m \geq m_{0}$ is {\em not} necessary. There are actually pathological sets $\Sigma$ whose normalized secant set has an {\em infinite upper box counting dimension} and for which some operators $M: \sH \to \bR^{m}$ {\em with only $m=1$} measurement satisfy the RIP \cite{Puy_2015}.
\end{remark}
\subsubsection{Some examples}

When given a random linear operator $A: \sH \to \bR^{m}$, one can leverage the above result to check whether $A$ satisfies the RIP with high probability. Alternatively, one can {\em construct} such an operator by pursuing the strategy described at the beginning of this section. We now need to choose the matrix $M \in \bR^{m \times d}$.
%
%
Examples of matrices $M \in \mathbb{R}^{m \times d}$ such that the operator $A = M P_H$ satisfies \eqref{eq:prob_bound_increment_1} and \eqref{eq:prob_bound_increment_2} are:
\begin{itemize}
\item matrices with independent random Gaussian entries with mean $0$ and variance $1/m$;
\item matrices whose entries are independent random Bernoulli variables $\pm 1/\sqrt{m}$;
\item matrices whose rows are independently drawn from the Euclidean sphere of radius $\sqrt{d/m}$ in $\mathbb{R}^d$ using the uniform distribution.
\end{itemize}
If $M$ is one of the above matrices  (or more generally a matrix with independent subgaussian rows), considering the orthogonally projected model set $\Sigma' = P_{H}\Sigma$, its normalized secant set $\sS' = \sS(\Sigma')$, and $s > \boxdim(\sS') = \boxdim(\sS)$, we have \cite{Puy_2015}: $M$ satisfies the RIP on $\Sigma' - \Sigma'$ with constant $\delta'< \delta_0$ with high probability provided
\begin{equation}
  m  \geq \frac{C'}{\delta_{0}^2} \, \max\left\{s \log\left(\frac{1}{\alpha_{\sS'}}\right),\log\left(\frac{6}{\xi}\right)\right\},
\end{equation}
where $C'$ is a constant that depends on the distribution of $M$.

\subsection{Summary}
To summarize, a generic strategy (a way to implement the strategy in Figure~\ref{fig:dimension_reduction}) to build a compressive sensing measurement operator for a set $\Sigma$ that has a normalized secant set $\sS$ of finite upper box-counting dimension is  : 

\begin{enumerate}
 \item Find a (potentially high-dimensional) finite-dimensional space $H$ whose orthogonal projection operator satisfies \eqref{eq:projection}. A generic construction ofsuch a space is presented in Section~\ref{sec:proj_finite_subspaces}.
 \item Compose this projection operator with a random projection operator $M$ (a random matrix) such that \eqref{eq:prob_bound_increment_1} and \eqref{eq:prob_bound_increment_2} holds.
\end{enumerate}

Now that we have described how a we can build operators preserving low complexity models, we can turn to the study of the performance of methods used to recover $x$ from $y$. 
  
 \section{Performance of regularizers for the recovery of low-dimensional models}\label{sec:recovering}
 
As they satisfy the RIP, the linear operators $A$ built with the technique just described in Section~\ref{sec:observing} preserve the low-dimensional model $\Sigma$ in the sense that stable reconstruction of vectors from $\Sigma$ is possible with the so-called ``ideal decoder''. Yet, this decoder is often intractable in practice as it involves possibly non-convex and/or non-smooth optimization. 
We now turn to general decoders, with an emphasis on convex decoders: minimization algorithms are well known for such decoders and they are often possible to implement with off-the-shelf algorithms, as in the classical cases of basis pursuit ($\ell^{1}$ norm minimization) or nuclear norm minimization.
 \subsection{Convex decoders and atomic norms}
 In the framework of minimization~\eqref{eq:robust_minimization}, it is interesting  to consider a particular class of convex functions: atomic norms with atoms included in the model set $\Sigma$~\cite{Chandrasekaran_2012}. Considering a set $\sA \subset \sH$, commonly called the set of {\em atoms}, the corresponding {\em atomic ``norm''} is built using the convex hull of $\sA$ .

\begin{definition}[Convex hull]
The convex hull of a set $\sA$ is: 
\begin{equation}
\tconv(\sA) \defin  \left\{\vx = \sum c_i \va_i: \va_i \in \sA, c_i \in \bRp, \sum c_i = 1\right\}
\end{equation}
\end{definition}

\begin{definition}[Atomic norm]
 The atomic ``norm'' induced by the set $\sA$ is defined as: 
\begin{equation}
 \|\vx\|_\sA \defin \inf \left\{ t \in \bRp:  \vx \in t\cdot\cl{ \tconv}(\sA) \right\}
\end{equation}
where $\cl{ \tconv}(\sA)$ is the closure of $\tconv(\sA)$ in $\sH$. The function $\|\vx\|_\sA$ is a convex gauge that is not always a norm. It is a norm if $\sA$ is symmetrical and bounded. We will keep the term atomic norm in the general case as an abuse of notation. This norm is finite only on the set
\begin{equation}
\sE(\sA) \defin \bRp \cdot \cl{\tconv}(\sA) = \{\vx = t\cdot\vy, t \in \bRp,\vy \in \cl{\tconv}(\sA)\} \subset \sH.
\end{equation}
It can be extended to $\sH$ by setting $\|\vx\|_\sA := +\infty$ if $\vx \notin \sE(\sA)$. 
\end{definition}

Atoms are often \emph{normalized}: a vector $u$ is \emph{normalized} if  $\|u\|_\sH=1$.

\begin{remark}
  Atomic norms are interesting because given any convex regularization function it is always possible to find an atomic norm that performs noiseless recovery better (in the sense that it permits recovery for more measurement operators $A$ \cite{Traonmilin_2016}). 
\end{remark}

\subsubsection{Classical examples of atomic norms}
As pointed out in \cite{Chandrasekaran_2012}, many well know norms used for low complexity recovery are atomic norms:
\begin{itemize}
\item  $\ell^1$ norm in $\bR^n$: $\sA$ is the set of canonical orthonormal basis vectors multiplied by a real scalar with modulus $1$, i.e. the normalized $1$-sparse vectors.
\item  Nuclear norm:  $\sA$ is the set of normalized rank one matrices.
\item Gauge generated by a finite polytope: $\sA$ is composed of the vertices of a polytope. 
\item Spectral norm: $\sA$ is the set of normalized orthogonal matrices.
\end{itemize}

\subsubsection{Group norms in levels}

For our running example, the model $\Sigma = \Sigma_{1} \times \ldots \Sigma_{J}$ associated to structured sparsity in levels, we consider a similar class of atomic norms: the group norms in levels. 

Given the subspace $\sH_{j}$ associated to the $j$-th level, $\sS_{j}(1) \subset \sH_{j}$ its unit sphere, $G_{j}$ its set of groups, and $\Sigma_{1,j}$ the associated $1$-group sparse model, consider the collection of atoms of the $j$-the level:
\begin{equation}\label{eq:DefKGroupSparseAtoms}
\sA_j \defin \Sigma_{1,j} \cap \sS_j(1).
\end{equation}
The corresponding  atomic norm is associated to the finite-dimensional space 
\[
\sE(\sA_j) = \linspan(\{\ve_{i}\}_{i \in \cup_{g \in G_j} })
\]
and simply given by
\begin{equation}\label{eq:DefKGroupSparseAtomicNorm}
\|\vx\|_{\sA_j} = 
\begin{cases}
\sum_{g\in G} \|\vx_g\|_\sH,\ &\vx \in \sE(\sA);\\
+\infty,\ & \vx \notin \sE(\sA)
\end{cases}
\end{equation}
The norm $\|\vx\|_{\sA_j}$ is called a {\em group norm}, a {\em structured norm} or a {\em mixed $\ell^1-\ell^2$ norm} \cite{Yuan_2006}. 

A natural regularizer for the \emph{structured sparsity in levels} model is defined as follows in $\sH_{1} \times \ldots \times \sH_{J}$: 
\begin{equation}\label{eq:DefWeightedStructuredNormLevels}
f_w: (\vx_1, \ldots \vx_J) \mapsto w_1\|\vx_1\|_{\sA_1}+\ldots+w_J\|\vx_J\|_{\sA_J}
\end{equation}
with weights $w_j>0$. We will show in the next sections that setting appropriately the weights in each level can yield recovery guarantees of various strengths.

\subsubsection{Atomic norm associated to a union of subspace model}

\begin{figure}[h]
\centering
\subfloat[$\|\cdot\|_1$]{\includegraphics[width=0.32\linewidth]{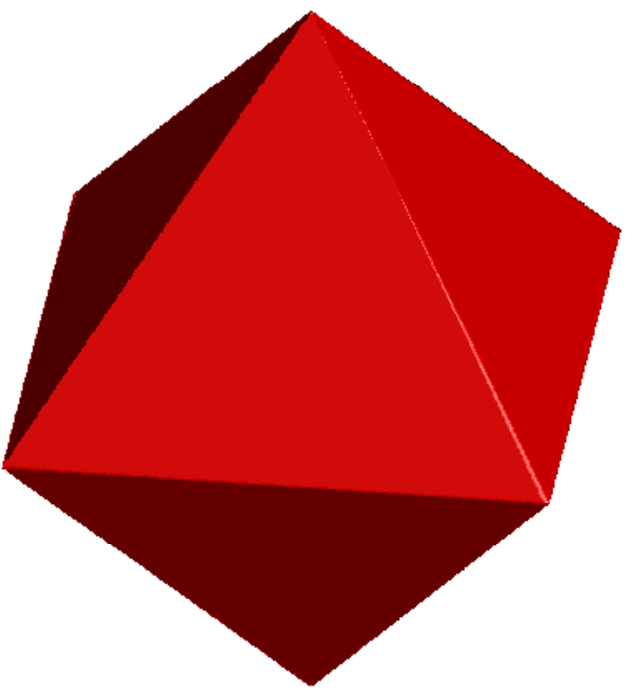}}\hspace{10mm}
\subfloat[$\|\cdot\|_\Sigma$]{\includegraphics[width=0.35\linewidth]{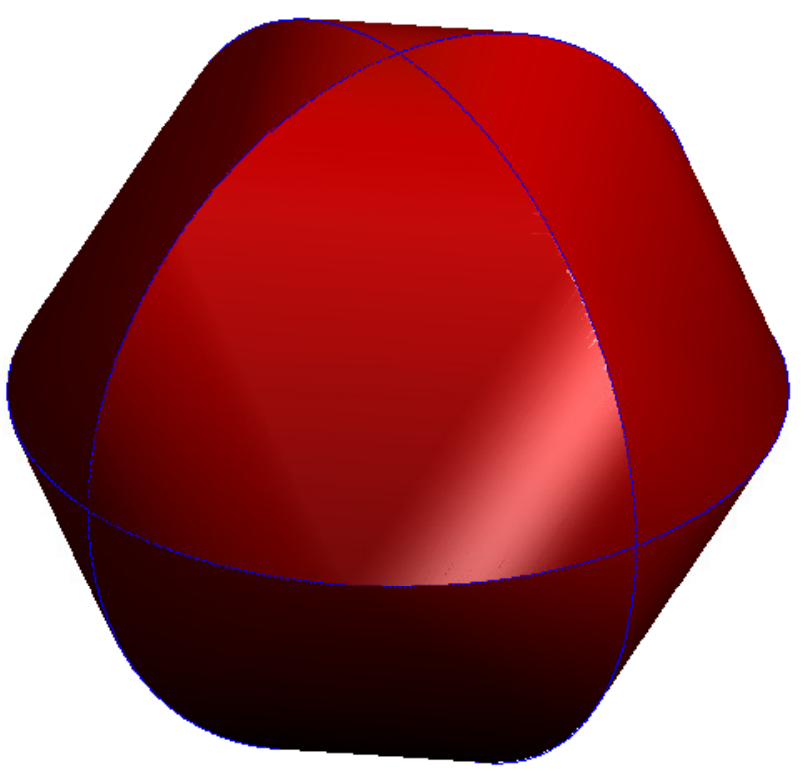}}
\caption{The unit ball of $\|\cdot\|_1$ (left) and the unit ball of $\|\cdot\|_\Sigma$ ($k$-support norm) for $\Sigma =\Sigma_2$ the set of 2-sparse vectors in 3D (right).}\label{fig:norm_sigma}
\end{figure}

Many classical model sets $\Sigma$ (the set of sparse vectors, the set of low-rank matrices, etc.) are homogeneous: if $x \in \Sigma$ then $\alpha x \in \Sigma$ for any scalar. As such they are (finite or infinite) unions of subspaces. Given any union of subspaces $\Sigma \subset \sH$, the norm associated to its {\em normalized atoms} 
\begin{equation}
\sA(\Sigma) \defin \Sigma \cap \sS(1)
\end{equation}
will be of particular interest for the RIP analysis described in the next sections. As a shorthand notation, we define
\begin{equation}
\|\cdot\|_\Sigma \defin \|\cdot\|_{\Sigma \cap \sS(1)}.
\end{equation}
This norm is sometimes useful as a regularizer to perform recovery (i.e. by choosing $f(z) = \|z\|_{\Sigma}$ in minimization~\eqref{eq:robust_minimization}). For the particular case where $\Sigma$ is the set of $k$-sparse vectors, $\|\cdot\|_{\Sigma}$ is know as the $k$-support norm \cite{Argyriou_2012}. It is known to yield stable recovery guarantees for certain $k$-sparse vectors \cite{Argyriou_2012}, however it has been shown that these results cannot be made uniform for all $k$-sparse vectors (and consequently similar negative results hold for structured sparsity in levels)~\cite{Traonmilin_2016}. We show in Figure~\ref{fig:norm_sigma} a representation of the $\ell^1$-norm and of the $k$-support norm $\|\cdot\|_\Sigma$ for $k=2$ in 3D ($\sH = \mathbb{R}^{3}$), which are two atomic norms induced by normalized atoms included in the model set $\Sigma$.

\subsection{Stable and robust recovery of unions of subspaces} 

 The main result from \cite{Traonmilin_2016} states that the stability of {\em any} decoder of the form \eqref{eq:robust_minimization} is guaranteed provided the linear operator $A$ satisfies a RIP on the secant set $\Sigma-\Sigma$ with a constant $\delta<\delta_\Sigma(f)$ holds, where $\delta_\Sigma(f)$ is a constant that depends only on the regularizer $f$ and the model set $\Sigma$ (we give and discuss the definition of $\delta_\Sigma(f)$ in Section~\ref{sec:def_RIP_const} below).
 
\subsubsection{Stable recovery in the presence of noise}
Elements of the model can be stably recovered \cite[Theorem 1.2]{Traonmilin_2016}:
\begin{theorem}[RIP condition for stable recovery of a union of subspaces]\label{th:robust_RIP_eucl}
Assume that $\Sigma$ is a union of subspaces. Then, for any continuous linear operator $A$ on $\sH$  that satisfies the RIP on the secant set $\Sigma-\Sigma$ with constant $\delta < \delta_\Sigma(f)$ we have: for all $\vx\in\Sigma$, $e \in \sF$ such that $\|\ve\|_\sF  \leq \epsilon$ (recall that $\epsilon$ is an \emph{estimation} of the noise level used as a parameter of the decoder), with $\vx^*$ the result of minimization~\eqref{eq:robust_minimization},
\begin{equation}
\| \vx^*-\vx\|_\sH \leq   C_{\Sigma}(f,\delta) \cdot (\|\ve\|_\sF  +\epsilon )
\end{equation}
where $C_{\Sigma}(f,\delta) < +\infty$. 
\end{theorem}
We refer the reader to \cite[Theorem 1.2]{Traonmilin_2016} for an explicit expression of $C_{\Sigma}(f,\delta)$. It is increasing with respect to the RIP constant $\delta$ : the worse the RIP constant is, the worse the stability constant is (see for example its expression for structured sparsity in levels in Theorem~\ref{th:inst_opt_block}).
\subsubsection{Robustness to modeling error}
Regarding robustness to modeling error, generic results often use the so-called $A$-norm \cite{Bourrier_2014} (not to be confused with the atomic norm: here the $A$ refers to the measurement operator) as an intermediate tool to measure the distance from a vector $x$ to the model set $\Sigma$. Given a constant $C$, the $A$-norm is defined by 
\begin{equation}
\label{eq:DefMNorm}
\|\cdot\|_{A,C} := C \cdot \|A\cdot\|_\sF + \|\cdot\|_\sH.
\end{equation}
It is more convenient to express robustness results with respect to a norm that does not depend on the measurement operator $A$. We provide here a robustness result where the modeling error with respect to the regularizer $f$ is used (this is more in line with the classical literature for $\ell^1$ minimization of nuclear norm minimization). 
Consider the (symmetrized) distance with respect to $f$: 
\begin{equation}
 d_{f}(\vx,\Sigma) = \underset{\tilde{\vx} \in \Sigma}{\inf} \frac{f(\vx-\tilde{\vx})+f(\tilde{\vx}-\vx)}{2}.
\end{equation}
When  $f$  is a positively homogeneous, non-negative and convex regularizer that bounds the $A$-norm, robustness with respect to $d_f$ also generally holds \cite[Theorem 3.2]{Traonmilin_2016}:
\begin{theorem}\label{th:inst_opt2}
Let $\Sigma$ be union of subspaces. Let $f$ be positively homogeneous, non-negative and convex with $f(x)<+\infty$ for $x\in \Sigma$. Consider a continuous linear operator $A$ satisfying the RIP on $\Sigma-\Sigma$ with constant $\delta<\delta_\Sigma(f)$, and a noise level $\|\ve\|_\sF \leq \epsilon$. Denote $C_{\Sigma}$ the constant from Theorem~\ref{th:robust_RIP_eucl}, and assume that for all $u \in \sH$, $\|u\|_{A,C_\Sigma}\leq  C_{f,A,\Sigma} \cdot f(u)$ for some  $C_{f,A,\Sigma} < \infty$. 
Then, for all $\vx \in \sH$, $\ve \in \sF$,  such that  $\| \ve \|_\sH \leq \eta \leq \epsilon$, any minimizer $\vx^*$ of~\eqref{eq:robust_minimization} satisfies
\begin{equation}
 \|\vx^*-\vx\|_\sH \leq  C_\Sigma \cdot (\|\ve\|_\sF  +\epsilon) + 2 C_{f,A,\Sigma} \cdot d_f(x_0,\Sigma).
\end{equation}
\end{theorem}

\begin{remark}
To apply this theorem we need $C_{f,A,\Sigma}<\infty$. This is the case for most classical examples (sparse recovery with $\ell^1$-norm, low-rank matrix recovery with the nuclear norm). It is also true for the case where $f$ being a convex gauge induced by a bounded closed convex set containing $0$ and $\sH$ is of finite dimension.
\end{remark}

\begin{remark}
Both Theorem~\ref{th:robust_RIP_eucl} and Theorem~\ref{th:inst_opt2} can be extended to the case where $\Sigma$ is a {\em cone} instead of a union of subspaces, with a  definition of $\delta_\Sigma(f)$ adapted compared to the one given later in Section~\ref{sec:def_RIP_const} (See Section~\ref{sec:generality}).
\end{remark}

\subsubsection{Example: the case of sparsity in levels}\label{sec:exrecovsplev}
Consider the model set $\Sigma$ corresponding to our running example of structured sparsity in levels, and choose as a regularizer the weighted atomic norm $f_{w}(\cdot)$ defined in~\eqref{eq:DefWeightedStructuredNormLevels}. One can show \cite[Theorem 4.1]{Traonmilin_2016} that $\delta_{\Sigma}(f_w) \geq \frac{1}{\sqrt{2}}$ for $J=1$ and 
\[
\delta_{\Sigma}(f_w) \geq \frac{1}{\sqrt{2+J \kappa_w^2}}
\]
for $J \geq 2$, where $\kappa_w := \max(w_j\sqrt{k_j})/\tmin(w_j\sqrt{k_j})$. In particular, for the particular weights $w_j = 1/\sqrt{k_j}$, we have $\delta_{\Sigma}(f_{w}) \geq \frac{1}{\sqrt{2+J}}$ for $J \geq 2$. 
 
In comparison, Ayaz et al. \cite{Ayaz_2014} gave a uniform recovery result with the mixed $\ell^1-\ell^2$-norm for structured compressed sensing under a RIP hypothesis. They showed that a RIP constant  $\delta < \sqrt{2}-1$ for vectors in the secant set guarantees the recovery of vectors from the model. The above result shows that the RIP constant of Ayaz et al. can be improved to $\frac{1}{\sqrt{2}}$. In~\cite{Adcock_2013b}, a model of sparsity in levels was introduced: it is in fact a structured sparsity in levels model with classical sparsity (each group is reduced to a single coordinate) in each level. In \cite{Bastounis_2015}, Bastounis et al. showed that when the model $\Sigma$ is sparsity in levels and   $f(\cdot) = \sum_j \|\cdot\|_{\sA_j} = \|\cdot\|_1$ (i.e., with weights $w_{j}=1$, in this case, $\kappa_w^{2}=\kappa_1^{2}$  is the maximum ratio of sparsity between levels), the RIP with constant $\delta =1/\sqrt{J(\kappa_1+0.25)^2 +1)}$ on $\Sigma-\Sigma$ guarantees recovery. This constant is improved to the constant $\delta_\Sigma(f_{w})\geq 1/\sqrt{2+J}$ when weighting the norm of each level with $w_j = 1/\sqrt{k_j}$. The above result further extends the work of Bastounis et al. to general structured sparsity. The following theorem \cite[Theorem 4.3]{Traonmilin_2016} summarizes the result with this optimal weighting:
\begin{theorem}\label{th:inst_opt_block}
Let $\Sigma$ be the model set associated to structured sparsity in levels, and consider $f=f_w$ as a regularizer, with the adapted weights $w_j=1/\sqrt{k_j}$. Suppose the continuous linear operator $A$ satisfies the RIP with constant $\delta < \delta_\Sigma(f)$ on the secant set $\Sigma-\Sigma$. Then for all $\vx\in \sH$,  $\ve \in \sF$ such that $\|\ve\|_\sF  \leq \epsilon$, and $\vx^*$ the result of minimization~\eqref{eq:robust_minimization}, we have
\begin{equation}
\| \vx^*-\vx\|_\sH \leq   C_{\Sigma}(f,\delta)(\|\ve\|_\sF +\epsilon) + D_{\Sigma}(f,\delta) \cdot d_{f}(\vx,\Sigma)
\end{equation}
where : 
\begin{itemize}
 \item For $J=1$, $\delta_0 = \frac{1}{\sqrt{2}}$, $C_{\Sigma}(f,\delta) \leq  \frac{2\sqrt{1+\delta}}{1-\delta\sqrt{2}}$ and $D_{\Sigma}(f,\delta)= 2(1+\sqrt{1+\delta}C_{\Sigma}(f,\delta))/\sqrt{k}$.
 \item For $J\geq 2$, $\delta_0 =\sqrt{\frac{1}{2+J}}$, $C_{\Sigma}(f,\delta) \leq  \frac{(1+\sqrt{1+J})\sqrt{1+\delta}}{1-\delta\sqrt{2+J}}$ and $D_{\Sigma}(f,\delta)= 2\sqrt{2}(1+\sqrt{1+\delta}C_{\Sigma}(f,\delta))$.
\end{itemize}
\end{theorem}

This result recovers classical guarantees with $\ell^1$ minimization for sparse recovery. 
Since $\delta_{\Sigma}(f) \geq 1/\sqrt{2+J}$ for $J\geq 2$, combining Theorem~\ref{th:inst_opt_block} for $\delta < 1/\sqrt{2+J}$ 
with results from Section~\ref{sec:observing} yields \cite{Traonmilin_2016} that
\[
  m \geq 
   O\left(J\sum_{j=1}^{J} \left(k_{j}d_{j} + k_j \log\left(\tfrac{3e |G_j|}{k_j }\right) \right)\right).
\]
subgaussian measurements are sufficient to guarantee stable and robust recovery with $f_w$, where $w_{j}=1/\sqrt{k_{j}}$. 
\begin{remark}
The factor $J$ might seem pessimistic, and we attribute its presence to the generality of the result. Should the structure of the observation matrix $A$ be taken into account, better results can be achieved. In fact, if $A$ is a block diagonal matrix where each block $A_j$ has size  $m_j \times n_j$, uniform recovery guarantees with the $\ell^1$-norm
hold if and only if uniform recovery holds on each block: this is possible as soon as each block $A_j$ of $A$ satisfies the RIP with some constant $\delta_j < \frac{1}{\sqrt{2}}$ on $\Sigma_j-\Sigma_j$, which is in turn exactly equivalent to the RIP with constant $\delta < \frac{1}{\sqrt{2}}$ on $\Sigma-\Sigma$.
\end{remark}

\begin{remark}
To make sense of Theorem~\ref{th:inst_opt_block} in the \emph{infinite-dimensional setting}, the domain where the regularizer $f$ is finite must be extended outside of $\sE(\Sigma)$ while keeping a finite  constant $D_{\Sigma}(f,\delta)$ . This can be done on a case-by-case basis when properties of $A$ and $f$ allow to conclude. For example, as Adcock and Hansen in \cite{Adcock_2013b}, consider the following setting: $\sH=\ell^2(\sN)$ with Hilbert basis $(e_i)_{i=1,+\infty}$. Consider $\Sigma$ a sparsity in levels model in $(e_1,..,e_N)$. Let $f=\|\cdot\|_1$. Then $f$ is an extension of the definition of $f_{w}$ in $\sE(\Sigma)$ to the whole space $\sH$ (with $w_j=1$ for all $j$). In \cite{Adcock_2013b}, the measurement operator $A$ is a collection of (Fourier) measurements that have \emph{a strong balancing property}. The important fact here is that this property requires $\|A^HA\|_{\infty} \leq C'$ where $\|\cdot\|_{\infty}$ is the maximum of the $\ell^\infty$-norms of the coefficients of $A^HA$ (where $A^H$ is the Hermitian conjugate of $A$). With such an hypothesis, for any $u \in \sH$, we have:  $\|Au\|_2^2 = | \ls u, A^HAu \rs| \leq \|A^H A u\|_{\infty} \|u\|_1\leq \|A^HA\|_{\infty} \|u\|_1 \|u\|_1 \leq C' \|u\|_1^2 $. Thus in this case the $A$-norm is bounded by the $\ell^1$-norm: $\|\cdot\|_{A,C} \leq (1+C \sqrt{C'})\|u\|_1$. 
\end{remark}
 \subsection{Definition and calculation  of $\delta_\Sigma(f)$}\label{sec:def_RIP_const}
    
 When $\Sigma$ is a union of subspaces, the sufficient RIP constant for recovery of elements of $\Sigma$ with $f$ is defined as 
\begin{equation}
 \delta_\Sigma(f) \defin \underset{\vz\in \sT_f(\Sigma) \setminus \{0\} }{\inf}\ \ \underset{\vx \in \Sigma}{\sup}\ \ \delta_{\Sigma}(\vx,\vz).
\end{equation}
where
\begin{eqnarray}
\delta_{\Sigma}(\vx,\vz)  &\defin&  \frac{-\re \ls \vx,\vz \rs}{\|\vx\|_\sH \sqrt{ \|\vx+\vz\|_\Sigma^2 -  \|\vx\|_\sH^2 - 2\re \ls \vx,\vz \rs}}.\label{eq:DefDeltaUoS}\\
\end{eqnarray}
 and $ \sT_f(\Sigma) $ is the set of descent vectors of $f$ at points of $\Sigma$: 
 \begin{equation}
 \sT_{f}(\Sigma) \defin \left\{ \vz \in \sH: \exists \vx \in \Sigma \; / \; f(\vx+\vz) \leq f(\vx) \right\}
 \end{equation}
 
 It is important to note that the constant $\delta_\Sigma(f)$ only depends on the geometry of $\Sigma$ and $f$. This constant measures the quality of $f$ as regularizer to recover elements of $\Sigma$ under a RIP assumption: the larger $\delta_\Sigma(f)$, the weaker the assumption on the linear operator $A$ to ensure stable and robust recovery in Theorem~\ref{th:inst_opt2}.  
 
To obtain concrete results, one needs to lower bound the above expression. As the supremum in the expression of $\delta_\Sigma(f)$ is a priori hard to compute explicitly, for $z\in  \sT_f(\Sigma)$ one can intuitively seek an element $\vx \in \Sigma$ that maximizes the correlation with $-z$ (i.e., such that $-\re \ls \vx , \vz \rs$ is maximized).
  For the model associated to structured sparsity in levels, with the regularizer $f=f_w$ and weights $w_j= 1/\sqrt{k_j}$, this consists in taking $x = -z_T$ where $T$ is the support such that $z_T \in \Sigma$ and $z_T$ concentrates the most energy of $z$. With such an $x$, denoting $z_{T_c} = z-z_{T}$, one can show that
  \begin{equation}
   \delta(-z_T,z)=  \frac{1}{\sqrt{ \frac{\|\vz_{T_c}\|_\Sigma^2}{\|\vz_T\|_\sH^2} + 1}}
  \end{equation}
Since $z\in \sT_f(\Sigma)$, one shows that the fact that $z_T$ concentrates the most energy implies that $f(z_{T_c})\leq f(z_T)$, which in turns allows one to conclude that $\|\vz_{T_c}\|_\Sigma^2/\|\vz_T\|_\sH^2 \leq 1+J$ is bounded using a control of $ \|\vz_{T_c}\|_\Sigma^2$ obtained by extending Cai's sparse decomposition of polytopes~\cite{Cai_2014}. This leads to the bound $\delta_\Sigma(f_{w})\geq 1/\sqrt{2+J}$ mentioned in Section~\ref{sec:exrecovsplev}.

\section{Generality of the whole framework}\label{sec:generality}

The proof of the existence of random linear maps that reduce dimension while satisfying the RIP is valid for any finite-dimensional model set  $\Sigma$ in any Hilbert space. The guarantees for convex decoders from Section~\ref{sec:recovering} allow to define a critical RIP value for any union of subspaces $\Sigma$ and any regularizer $f$ (for some pairs this may yield $\delta_{\Sigma}(f) =0$, e,g, the right hand side of Figure~\ref{fig:norm_sigma}). Overall, the compressive sensing framework described in this chapter is thus very general and we give here an overview of examples where it applies.

\subsection{A flexible way to guarantee recovery}
The following list summarizes the results of the combined framework of Section~\ref{sec:observing} and~\ref{sec:recovering} for classical pairs of model $\Sigma$ and regularizer $f$. It states the model $\Sigma$, the considered regularizer $f$, a lower bound on our sufficient RIP constant $\delta_{\Sigma}(f)$, and a sufficient number of (random subgaussian) measurements $m$ to guarantee recovery using the construction from Section~\ref{sec:observing}.

\begin{itemize}
 \item  $\Sigma =$ Linear subspace of dimension $n$, $f =$ indicator function $\iota_{\Sigma}$ or $\|\cdot\|_\Sigma$.  \\
 \mytextbullet\ $\delta_\Sigma(f)=1$: this sufficient RIP constant was already known, e.g.  \cite{Bourrier_2014}. \\
 \mytextbullet\ Sufficient number of measurements:  $m \gtrsim n$.
 \item  $\Sigma =$ $k$-sparse vectors in dimension $n$, $f =$  $\ell^1$-norm. \\
  \mytextbullet\ $\delta_\Sigma(f)\geq 1/\sqrt{2}$. This is the sharp RIP constant of Cai et al. \cite{Cai_2014} (sharpness will be discussed in the next section). \\
 \mytextbullet\ Sufficient number of measurements:  $m \gtrsim k log(n)$.  
 \item   $\Sigma =$ Matrices of rank lower than $r$ in dimension $n \times n$,  $f =$ nuclear norm. \\
 \mytextbullet\ $\delta_\Sigma(f)\geq 1/\sqrt{2}$. This is also the sharp RIP constant of Cai et al. \cite{Cai_2014}.\\
 \mytextbullet\ Sufficient number of measurements:  $m \gtrsim  r n$.   
 \item   $\Sigma =$ Finite union of $k$ 1D-half-spaces with coherence $\mu(\Sigma)$, $f = \|\cdot\|_\Sigma$. \\
 \mytextbullet\ $\delta_\Sigma(f)\geq\frac{2(1-\mu(\Sigma)) }{ 3+2\mu(\Sigma) }$. \\ 
 \mytextbullet\ Sufficient number of measurements:  $m\gtrsim \log(k)/\delta_{\Sigma}(f)^2$.  
 \item  $\Sigma =$  Permutation matrices  of dimension $n \times n$, $f = \|\cdot\|_\Sigma$.\\
 \mytextbullet\   $\delta_\Sigma(f) \geq\frac{2}{3}$.  \\
 \mytextbullet\ Sufficient number of measurements:  $m \gtrsim n\log(n)$.  
\end{itemize}

Combining the formalism of \cite{Puy_2015} with that of \cite{Traonmilin_2016}, these models can also be considered in an infinite-dimensional space $\sH$, which is convenient to handle analog compressive sensing scenarios.  Stable recovery guarantees are still valid in this infinite-dimensional setting. For robustness, one must make sure that the constant $C_{f,A,\Sigma}$ is finite. For convex $f$, this might need some further assumptions on the behaviour of $f$ and $A$ outside of the  space $\sE(\Sigma)$ (the subspace spanned by $\Sigma$) as mentionned in Section~\ref{sec:exrecovsplev}. 

\subsection{Uniform vs non-uniform recovery guarantees}
The framework described in this chapter focuses on {\em uniform} recovery guarantees for arbitrary linear operators.
Another trend of general framework for compressive sensing focuses on non-uniform guarantees for Gaussian observations. In particular, Chandrasekaran et al~\cite{Chandrasekaran_2012} studied the general non-uniform recovery from Gaussian observations with atomic norms. In this case, the goal is to show that, for any element $x$ of the model, atomic norm minimization will recover $x$ from $Ax$ with high probability on the draw of $A$. In contrast, in the framework presented in this chapter, we established conditions so that (with high probability) {\em the same} linear operator $A$ (i.e., a particular draw of a random operator) allows to stably and robustly recover {\em all elements of the model} with arbitrary regularizers. Moreover, these results are proved for general random matrices (typically, subgaussian matrices).

\subsection{Extensions}

The guarantees for convex decoders for unions of subspaces from Section~\ref{sec:recovering}  have further been extended to the case where the model set is a cone (a positively homogeneous sets) \cite[Theorem 3.1]{Traonmilin_2016}. This covers  models such as (subsets of) the cone of positive semi-definite matrices, or that of non-negative matrices.

Beyond the pure Hilbert norm setting described in this chapter, the generalized definition of the RIP from equation~\eqref{eq:RIP3} or its further generalizations to arbitrary norms in $\sH$ can be used \cite{Puy_2015} to establish dimension reduction results for structured acquisition. An example is the use of random rank one projections (which are a subset of sub-exponential random matrices) \cite{Cai_2015}, which offer a computationally efficient way to gather linear observations of a matrix, thus making them an interesting observation method for algorithmic purposes in the low rank matrix recovery problem.  
While the RIP constant $\delta_\Sigma(f)$ has not be extended to such settings yet, such developments seem accessible.

 
\subsection{Sharpness of results?}
In \cite{Puy_2015} the finite dimension of the normalized secant set allows one to conclude on the possibilities in terms of dimension reduction. Only a number of measurements of the order of the dimension is sufficient. However, this hypothesis is not necessary. It is possible to find a model $\Sigma$ whose normalized secant set $\sS$ has infinite upper box counting dimension such that there exists a measurement operator with the RIP on $\sS$. Hence a weaker necessary and sufficient condition on the ``dimension'' of $\sS$ could exist to guarantee the existence of measurements operators with stable dimension reduction capabilities.
     
In terms of recovery for arbitrary regularizers, it has been shown that a sufficient RIP constant can be provided. For classical families of models and regularizers (sparse recovery with the $\ell^1$ norm and low-rank matrix recovery with the nuclear norm), as well as for structured sparsity and the associated group norm the constant $\delta_\Sigma(f)$ is sharp in the following way: we know that there exist RIP matrices with constant arbitrarily close to $\delta_\Sigma(f)$ (here $1/\sqrt{2}$) which do not permit uniform recovery \cite{Davies_2009,Cai_2015} for some dimension of $\sH$ and some sparsity $k$ (or rank $r$). Considering sparsity in levels, we observe that $\delta_\Sigma(f)$ complies with the necessary dependency on the ratios of sparsity between levels and the number of levels $J$ \cite{Bastounis_2015}.  These sharpness results all consider \emph{families} of models and regularizers: it is a worst case sharpness among these families of regularizer. However, one can consider the question of strong sharpness: for a given model $\Sigma$ and regularizer $f$, what is the biggest RIP constant sufficient to guarantee recovery?

\subsection{New Frontiers: Super-resolution and compressive learning}
 
Much of the algorithmic and mathematical techniques revolving around the notion of sparsity in the context of inverse problems and compressive sensing have been developed with finite-dimensional models, involving e.g. a discretization of the time domain, or of the frequency domain. However, the physical phenomena underlying the acquisition of modern data from the analogue world are rather intrinsically continuous \cite{Vetterli_2002}. The generic framework for inverse problems and dimension reduction presented in this chapter is directly set up in an arbitrary Hilbert space setting, and as such it opens new perspectives for handling the analogue nature of many problems.  

Super-resolution is one such problem. In super-resolution, one aims at recovering spikes combinations of few spikes from their low pass observation. While spikes are usually modeled with Dirac measures, which can be considered as belonging to certain Banach spaces of measures (e.g., equiped with the total variation norm), one way of bringing super-resolution close to the content of this chapter is to consider a kernel metric, which will bring a Hilbert structure to such Banach spaces~\cite{Gretton_2013}. Intuitively, this amounts to choosing a high resolution at which we will measure energy in the signal space. In this context, all the results on recovery guarantees and dimension reduction hold. Several questions remain standing: is it possible to find a sufficient RIP constant $\delta_\Sigma(f)$ that also holds in this context? Do usual models in Banach spaces have a normalized secant set with finite dimension? With the work of \cite{Candes_2013,Castro_2017,Duval_2015}, we already know that low pass filtering allows to recover spikes up to some resolution with a convex decoder. 

Another related problem is compressive learning. In \cite{Bourrier_2013, Keriven_2015} it is shown empirically that Gaussian mixtures can be recovered from a so-called {\em sketch} of the data, which can be considered as random Fourier measurements of their probability density. Recent works suggests that for an an appropriately chosen kernel metric, the secant set of sufficiently separated mixtures of Diracs is of finite dimension for appropriately chosen kernel metric~\cite{Keriven_2016b}. It is then possible to guarantee the success of the ideal decoder with random observations. Practical results have been obtained using a greedy heuristic approach to the problem~\cite{Keriven_2016}. These results seem to indicate a possible generalization of the theory of dimension reduction and convex recovery to these problems.

\section*{Acknowledgment}
This work was supported in part by the European Research Council PLEASE project (ERC-StG-2011-277906) and the European Research Council C-SENSE project (ERC-ADG-2015-694888).
M. E. Davies would like to acknowledge the support of EPSRC grant EP/J015180/1.

\bibliographystyle{abbrv}
  \bibliography{chapter_bib}

\end{document}